\documentclass[10pt,letterpaper]{article}
\usepackage{opex3}

\begin{document}

\title{Stroke saturation on a MEMS deformable mirror for woofer-tweeter adaptive optics}

\author{Katie Morzinski,$^{\textrm{1,2}\ast}$ Bruce Macintosh,$^\textrm{1,3}$ Donald Gavel,$^\textrm{1,2}$ and Daren Dillon$^\textrm{1,2}$}

\address{$^\textrm{1}$ National Science Foundation Center for Adaptive Optics}
\address{$^\textrm{2}$ UCO/Lick Observatory, 1156 High St., University of California, Santa Cruz, CA, 95064, USA}
\address{$^\textrm{3}$ Lawrence Livermore National Laboratory, 7000 East Ave., Livermore, CA 94550, USA}
\email{ktmorz@ucolick.org} 



\begin{abstract}
High-contrast imaging of extrasolar planet candidates around a main-sequence star has recently been realized from the ground using current adaptive optics (AO) systems.
Advancing such observations will be a task for the Gemini Planet Imager, an upcoming ``extreme'' AO instrument.
High-order ``tweeter'' and low-order ``woofer'' deformable mirrors (DMs) will supply a \textgreater90\%-Strehl correction, a specialized coronagraph will suppress the stellar flux, and any planets can then be imaged in the ``dark hole'' region.
Residual wavefront error scatters light into the DM-controlled dark hole, making planets difficult to image above the noise.
It is crucial in this regard that the high-density tweeter, a micro-electrical mechanical systems (MEMS) DM, have sufficient stroke to deform to the shapes required by atmospheric turbulence.
Laboratory experiments were conducted to determine the rate and circumstance of saturation, i.e. stroke insufficiency.
A 1024-actuator 1.5-$\mu$m-stroke MEMS device was empirically tested with software Kolmogorov-turbulence screens of r$_0=$10--15~cm.
The MEMS when solitary suffered saturation $\sim$4\% of the time.
Simulating a woofer DM with $\sim$5--10 actuators across a 5-m primary mitigated MEMS saturation occurrence to a fraction of a percent.
While no adjacent actuators were saturated at opposing positions, mid-to-high-spatial-frequency stroke did saturate more frequently than expected, implying that correlations through the influence functions are important.
Analytical models underpredict the stroke requirements, so empirical studies are important.
\end{abstract}





\section{Introduction}

Direct imaging of extrasolar planet candidates around main-sequence stars has recently been realized\cite{marois2008,kalas2008,lagrange2008} employing current adaptive optics (AO) systems and the angular differential imaging technique\cite{marois2006}.
Such high-contrast observations with ground-based telescopes require careful control of the incident wavefront, eliminating noise so that faint sources can be detected.
Designed specifically for high-contrast imaging, the Gemini Planet Imager (GPI)\cite{macintosh2008, kalas_gpi} will use an ``extreme'' adaptive optics (ExAO) system to flatten the incident wavefront.

Key to this ExAO system is the high-order deformable mirror (DM), a polysilicon micro-electrical mechanical systems (MEMS) DM developed by Boston Micromachines Corporation\cite{cornelissen2008}.
Consisting of thousands of individually addressable actuators in a coin-sized space, MEMS DMs are a low-cost approach to high-order wavefront correction;
however, MEMS devices exhibit less mechanical stroke than conventional piezo-actuated DMs.

Stroke saturates when the desired DM shape at a given actuator is beyond that actuator's range of motion.
Saturation degrades the wavefront correction and scatters light into the region that should remain dark for detection of faint sources.
This paper evaluates avoiding stroke saturation through addition of a second low-order, high-stroke DM in tandem with the MEMS DM.

\section{Background}

MEMS devices were chosen as the high-order wavefront corrector for GPI because of their cost-effectiveness, high-actuator density, and agreeable performance.
The mirrors can be batch-fabricated using a silicon foundry approach similar to that used to make integrated circuits, for a cost an order of magnitude cheaper per actuator than that of traditional piezoelectric DMs.
MEMS technology is scalable to the actuator count required for ExAO high-contrast imaging.

\begin{figure}[htbp]
\centering\includegraphics[width=9cm]{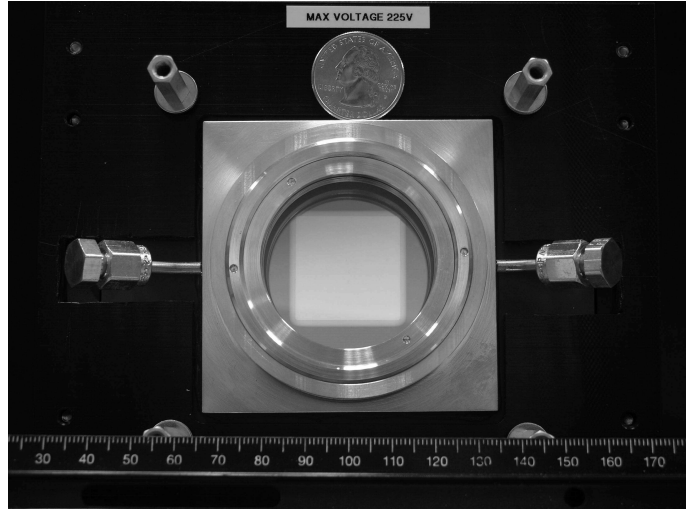}
	\caption{
		\label{fig:4kmems}
		4096-element MEMS deformable mirror in the Laboratory for Adaptive Optics.
		Center square is the reflective surface.
		The DM is packaged on a ceramic chip carrier
		and enclosed within a refillable dry-nitrogen-gas chamber
		behind a pressure window.
		}
\end{figure}

A series of prototype MEMS devices (two 144-element, ten 1024-element, and two 4096-element DMs) has been thoroughly tested at the University of California at Santa Cruz's Laboratory for Adaptive Optics (LAO).
At the sub-nanometer level, hysteresis is negligible\cite{morzinski2008a} and MEMS devices are temporally stable and precisely positionable\cite{morzinski2006}.
A MEMS DM has been flattened to 0.54~nm~rms in the control band\cite{evans2006b} and has achieved $10^{-6}$ contrast at the far field\cite{evans2005}.
An AO system using a 144-actuator MEMS DM has been successfully tested on-sky at Lick Observatory\cite{gavel2008a,gavel2008b}.
Figure~\ref{fig:4kmems} shows the LAO's first 4096-actuator DM, called an ``engineering-grade'' device because it is useable for testing and characterizing but is not fully functional across the mirror as required for GPI's ``science-grade'' MEMS.

Residual low-frequency wavefront error in the GPI optical train will scatter star light into the region of the focal plane that should be kept dark in order to detect planets.
This is often referred to as the ``dark hole'' region, and its size is set by the smallest-spatial-frequency control band, i.e. the spacing of the MEMS actuators in the Fourier domain\cite{malbet1995}.
Actuators on the deformable mirror that are broken, dead, or stuck scatter light into the dark hole by virtue of their influence functions producing a wide PSF in the far-field image.
Saturated actuators are problematic for high-contrast imaging because, optically, insufficient stroke has the same effect as a broken, dead, or stuck actuator.

\begin{figure}[htbp]
	\centering
  \unitlength1cm
  \begin{minipage}[t]{6cm}
    \begin{picture}(6,6)
      \includegraphics[width=6cm,height=6cm]{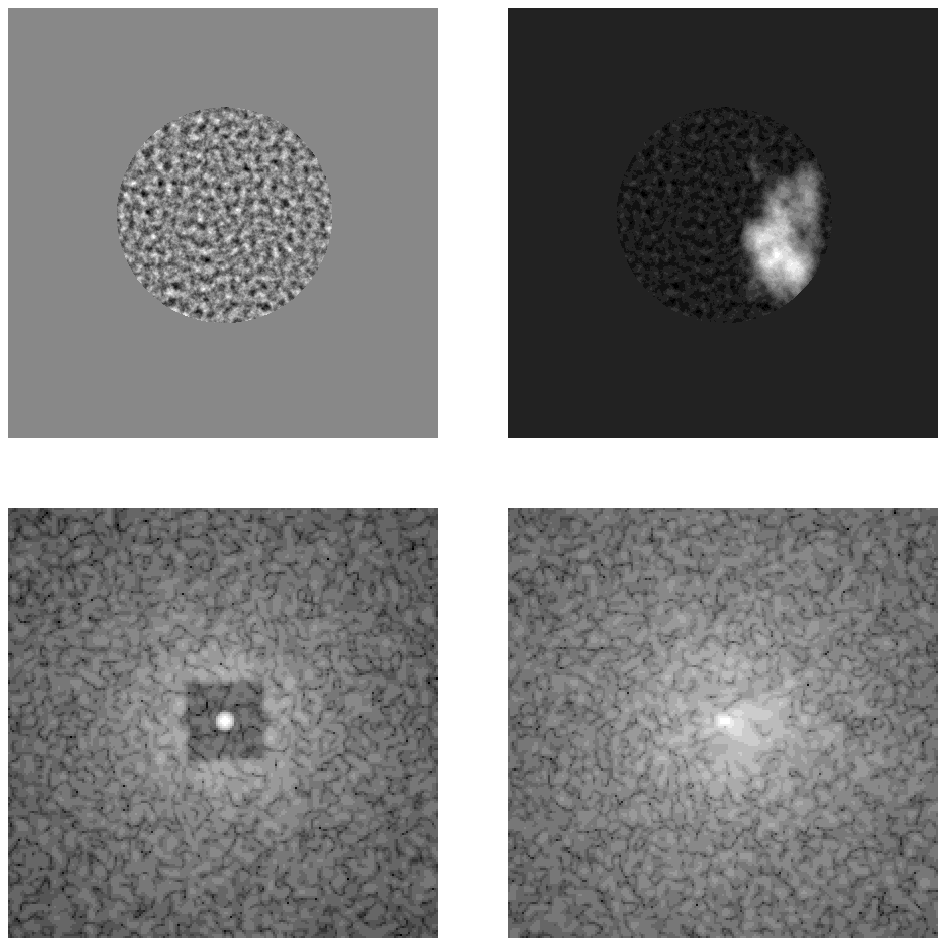}
    \end{picture}
  \end{minipage}
  \begin{minipage}[t]{7cm}
    \begin{picture}(7,5)
      \includegraphics[width=5cm,height=7cm,angle=90]{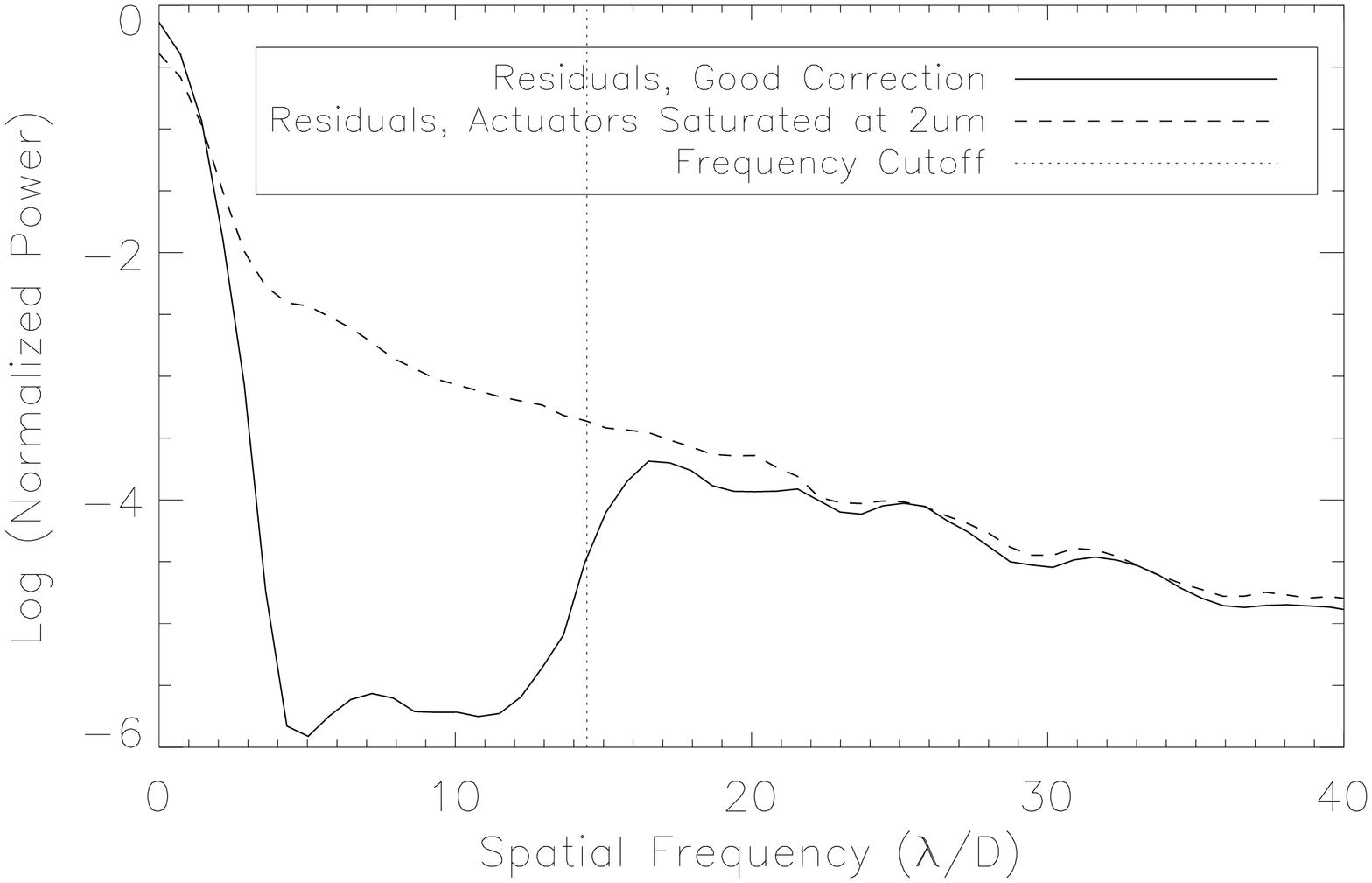}
	\end{picture}
  \end{minipage}
\caption{\label{fig:farfield}
		Illustration of the consequences of stroke saturation.
		\textit{Left top:} High-pass-filtered Kolmogorov phase screen, linear scale,
		and its simulated far-field image (\textit{left bottom}), log scale.
		\textit{Center top:} Exaggerated simulation of saturation with clipped Kolmogorov phase screen,
		linear scale,
		and its simulated far-field image \textit{(center bottom)}, log scale.
		\textit{Right:} Radial averages of the simulated far-field images.
		Solid line: High-Strehl AO correction.
		Dashed line: Stroke saturated above $2~\mu$m.
	}
\end{figure}

The following example (Fig.~\ref{fig:farfield}) shows the effects of actuator saturation.
A high-Strehl AO correction is effected by high-pass-filtering
a Kolmogorov phase screen (left top, linear scale), with the corresponding image at the far field (left bottom, log scale).
The far-field image is generated by the Fraunhofer approximation, propagating light from the phase-aberrated pupil plane (the DM) with an apodized transmission function\cite{soummer2005,thomas2008} to the focal plane.
If the phase measured at the MEMS plane is $\phi$ and the apodizer function is $A$, then the intensity $I$ at the far-field is simulated by calculating
\begin{equation}
I = \left|~\mathrm{FT}\left[{A~ \mathrm{exp}\left(\frac{2\pi}{\lambda}i\phi\right)}\right]\right|^2
\end{equation}
where $\lambda$ is the wavelength of the monochromatic light and $\mathrm{FT}$ means to take the Fourier transform.
The second case (center) is an exaggerated illustration of actuator saturation.
A low-pass-filtered Kolmogorov phase screen is clipped such that all values above $2~\mu$m are set to $2~\mu$m.
The residuals are displayed in linear scale (center top), and below that is the corresponding simulated far-field image in log scale.
At right, radial averages of the two far-field images show the dramatic contrast achieved in the dark hole at low spatial frequencies when there is a high-Strehl AO correction (solid line).
Therefore, stroke saturation (dashed line) cannot be tolerated since it obscures the dark hole and any planets located there.

The statistics of Kolmogorov turbulence determine the stroke required of the wavefront corrector, whereas physical constraints on the DM determine its stroke capabilities.
Variables that affect stroke capabilities include actuator design, size and spacing of actuators (i.e. pitch), and maximum voltage applied to the actuator.
Due to the small (340--400-$\mu$m) pitch and the continuous facesheet on the MEMS, bending stresses limit high-spatial-frequency stroke on current devices to $1.2~\mu$m surface.
The capacitor gap limits low-spatial-frequency stroke to $3.4~\mu$m surface.
Furthermore, stroke may also be drawn on to remove inherent curvature of the device and for tip/tilt residuals.

Previous LAO articles reported on MEMS stroke as a function of spatial frequency\cite{morzinski2006,morzinski2008b,norton2009}.
Three MEMS devices have been tested and the GPI MEMS has been specified for stroke at low and high spatial frequencies; Table~\ref{tab:spatfreq} compiles the results.
Norton \textit{et al.}\cite{norton2009} give more details of the ``engineering-grade'' 4k-MEMS characterization.
The GPI ``science-grade'' 4k-MEMS has not yet been delivered, but its specifications are given.

\begin{table}[htbp]
  \centering
  \begin{tabular}{cccccc}
    \hline
  MEMS & Actuator & Pitch & Max. Volt. & Low-freq.~stroke & High-freq.~stroke \\
    Device & Count & [$\mu$m] & Applied & [$\mu$m surface] & [$\mu$m surface] \\
    \hline
    \multicolumn{6}{l}{\textit{Empirical Measurements}} \\
    \hline
    W107\#X & 1024 & 340 & 160 & 1.0 & 0.2 \\
    W107\#X$^\dagger$ & 1024 & 340 & 200 & 1.5 & 0.2 \\
    W95\#39 & 1024 & 400 & 200 & 3.2 & 0.9 \\
    4kEng\#1 & 4096 & 400 & 225 & 3.4 & 1.2 \\
    \hline
    \multicolumn{6}{l}{\textit{Specifications}} \\
    \hline
    GPI 4k & 4096 & 400 & 260 & 4.0 & 1.0 \\
    \hline
  \end{tabular}
  \caption{
  Stroke measurements for a variety of MEMS DMs in the Laboratory for Adaptive Optics\cite{morzinski2008b}.
  $^\dagger$W107\#X at a maximum voltage of 200~V is the same MEMS device used for the
  Kolmogorov saturation experiments described in Section~\ref{sec:experiment}.
  }
  \label{tab:spatfreq}
\end{table}

\begin{figure}[htbp]
	\centering
  \unitlength1cm
      \includegraphics[width=9cm,angle=90]{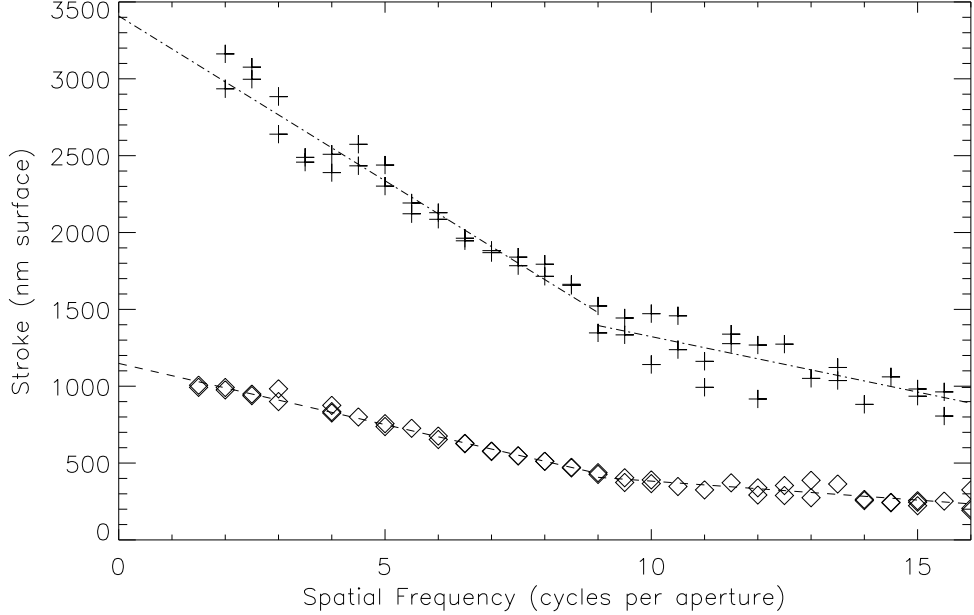}
\caption{\label{fig:spatfreq}
		LAO testing of stroke as a function of spatial frequency for two MEMS deformable mirrors.
		Diamonds:
		Device W107\#X, pitch $340~\mu$m, voltage range 0--160~V.
		Plus-signs:
		Device W95\#39, pitch $400~\mu$m, voltage range 0--200~V.
		The ``knee'' is located around 9 cycles per aperture for both devices.
		Note that a DM having sinc influence functions that fall to zero one actuator away
		would produce a horizontal line on this plot, as would a segmented mirror.
	}
\end{figure}

To measure MEMS stroke at a range of spatial frequencies, one-dimensional sinusoids of varying period were applied to two 1024-actuator DMs, and the resulting stroke was measured in each case\cite{morzinski2006,morzinski2008b}.
The results are plotted in Fig.~\ref{fig:spatfreq} for the devices W107\#X (pitch $340~\mu$m and tested on the voltage range 0--160~V) and W95\#39 (pitch $400~\mu$m and tested on the voltage range 0--200~V).
MEMS stroke decreases with increasing spatial frequency, whereas a horizontal line on Fig.~\ref{fig:spatfreq} would be produced by both a segmented mirror and a DM with actuators having narrow sinc influence functions
(with the first zero occurring at a spacing of one actuator away).
This departure from the narrow-sinc case indicates that the broad MEMS influence functions\cite{severson2006inffxn} have the effect of reducing the stroke as spatial frequency increases.

Rather than plotting a single line with each data series on Fig.~\ref{fig:spatfreq}, two line segments joined by a ``knee'' fit each measurement series better.
The knee in both cases is located around 9 cycles per aperture, a 3.6-actuator period, while the influence function measured for a 340-$\mu$m-pitch MEMS falls to 4\% at 2 actuators away\cite{severson2006inffxn}, about half this period.
This implies that the high-spatial-frequency stroke is controlled locally by neighboring-actuators' influence functions, whereas the low-spatial-frequency stroke is in a distinct whole-MEMS regime.

\section{Theory}

In a woofer-tweeter architecture, two DMs are arranged in series and both are conjugate to the same optical plane, usually that of the telescope pupil.
The low-actuator-count, high-stroke DM is termed the ``woofer,'' and the high-actuator-count, low-stroke DM is the ``tweeter,'' a MEMS in the case of GPI.
As a consequence of the Kolmogorov turbulence spectrum, atmospheric turbulence strength decreases with spatial frequency\cite{tatarski1961}, making the woofer-tweeter architecture feasible.
The woofer DM corrects the higher-stroke, lower-spatial-frequency aberrations while the tweeter corrects the lower-stroke, higher-spatial-frequency aberrations.

From Kolmogorov turbulence theory one can calculate the expected stroke requirement for a deformable mirror at a given telescope\cite{hardy1989}.
At Gemini South, the planned location of GPI, the primary is $D=8$~m and a typical Fried parameter is r$_0=14$~cm.
The power spectral density $\Phi$ of Kolmogorov turbulence goes as $\Phi(\kappa) \propto \kappa^{-5/3}$, where $\kappa$ is the wave number.
The formula for fitting error in radians $\sigma_{F}$ as a function of r$_0$ and the actuator spacing $d$ is
\begin{equation}
\sigma_{F}^{2} = \alpha_{F} \left(\frac{d}{r_{0}}\right)^{5/3},
\label{eqn:fittingerror}
\end{equation}
where $\alpha_{F}$ is a constant that depends on the influence function for the deformable mirror in question\cite{noll1976,hudgin1977}.

For a continuous facesheet DM, $\alpha_F = 0.14$ for circular segments with piston and tip/tilt actuation capabilities\cite{hardy1998}.
Let us take this value for $\alpha_F$, put the primary diameter $D=d=8$~m, r$_0=14$~cm, and $\lambda=500$~nm.
To calculate the total stroke requirement, we convert Equation~\ref{eqn:fittingerror} from square radians to microns surface, and multiply by 5 to obtain a $\pm5\sigma$ correction:
\begin{equation}
\label{eqn:strokerequirement}
\sigma_{F} = (\pm5) \left(\frac{1}{2}\right) \left(\frac{\lambda}{2\pi}\right) \sqrt{\alpha_F} \left(\frac{D}{r_{0}}\right)^{5/6} = \pm2.17~\mu\mathrm{m~surface} = 4.3~\mu\mathrm{m~surface~P-V.}
\end{equation}
The factor of $1/2$ converts from phase to surface and $\lambda/(2\pi)$ converts from radians to microns.
However, $4.3~\mu$m surface is the stroke requirement at the center of the pupil, and due to factors including non-stationarity of phase\cite{conan2008} (which adds 20\% at the edges of the pupil), inherent curvature of the MEMS (on the order of a few $\mu$m for current devices), and tip/tilt residuals, up to $\sim10~\mu$m stroke is desired\cite{lavigne2008}.
From Table~\ref{tab:spatfreq} we see that the 4k-MEMS devices do not get as much as 10$~\mu$m stroke, so it was determined that a woofer DM was needed.

Equation~\ref{eqn:fittingerror} can also be used to calculate the residuals after applying a woofer, which gives the stroke requirement for the tweeter in microns surface:
\begin{equation}
\sigma_F = \left(\frac{1}{2}\right) \left(\frac{\lambda}{2\pi}\right) \sqrt{\alpha_F} \left(\frac{d_{woof}}{r_{0}}\right)^{5/6},
\label{eqn:tweeter}
\end{equation}
where d$_{woof}$ is the woofer actuator spacing.
In this case the $\alpha_{F}$ parameter becomes 0.28---appropriate for the influence function of a continuous facesheet DM\cite{hardy1998}.

\begin{figure}[htbp]
\centering\includegraphics[width=9cm,angle=90]{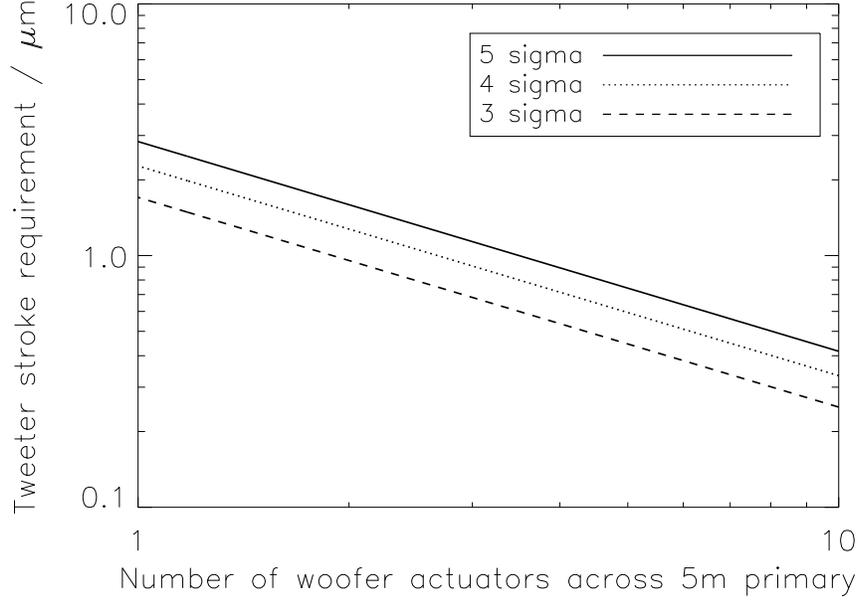}
	\caption{
		\label{fig:tweeterthy}
		Tweeter stroke required as a function of woofer degrees of freedom
		for r$_0=15$~cm, based on Equation~\ref{eqn:tweeter}.
		}
\end{figure}

Varying d$_{woof}$ in Equation~\ref{eqn:tweeter} and multiplying the final result by $\pm$3--5 to get the 3--5-sigma stroke requirement, the tweeter stroke required for each woofer degree of freedom is plotted in Fig.~\ref{fig:tweeterthy}.
Note that the stroke requirement quoted above of $10~\mu$m is for a 8-m pupil and r$_0=14$~cm
while Fig.~\ref{fig:tweeterthy} is for a 5-m pupil and r$_0=15$~cm.
For the parameters in Fig.~\ref{fig:tweeterthy},
using a woofer mitigates the tweeter stroke requirement to a more manageable $\sim$1$\mu$m surface.
Total stroke required rises as $D^\frac{5}{6}$ with primary diameter (i.e. Equation~\ref{eqn:strokerequirement}),
which increases the requirement on the woofer but not the tweeter.
Rather, tweeter stroke is a function of the woofer pitch, not the primary diameter.
Therefore, experiments simulating a 5-m pupil are still valid for the 8-m pupil case in regards to the tweeter.
Laboratory experiments follow to verify empirically the ability of the MEMS to form Kolmogorov atmosphere shapes, using a woofer to avoid saturation.

\section{Experiment}
\label{sec:experiment}

We tested MEMS device W107\#X for stroke saturation when being deformed to correct for Kolmogorov turbulence.
W107\#X has a 340-$\mu$m pitch and was applied with a maximum voltage of 200~V.
From Table~\ref{tab:spatfreq}, we see its low-order stroke is 1.5~$\mu$m surface and its high-order stroke is 0.2~$\mu$m surface.
While this DM does not have as much stroke as the final GPI 4k-MEMS will, the experiments are useful in determining whether inter-actuator or peak-to-valley stroke causes more saturation events.

\begin{figure}[htbp]
\centering\includegraphics[height=11cm]{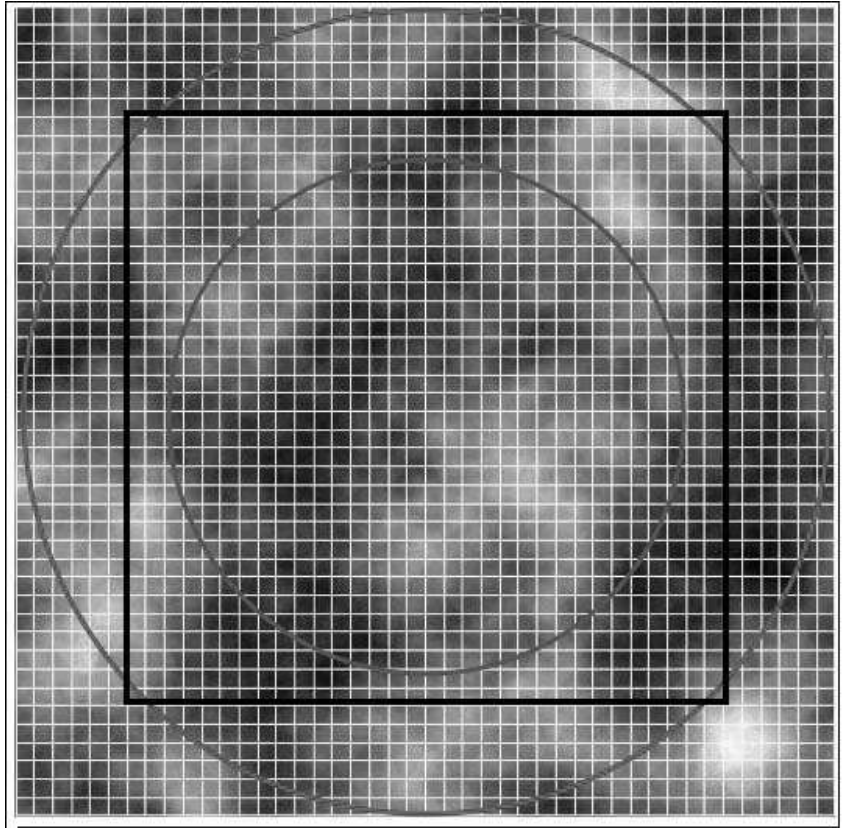}
	\caption{
		\label{fig:experimentscaling}
		To-scale mapping of 8m Gemini pupil (outer gray circle), 5m testbed pupil (inner gray circle),
		44x44-actuator array for GPI (full grid), and 32x32-actuator array of MEMS
		in testbed (black).  Fine white grid delineates individual MEMS actuators.
		}
\end{figure}

Device W107\#X has 1024 actuators (32 across) rather than the 44-actuators-across GPI will use.
Thus, because the high spatial frequencies on the MEMS were of more interest than the low, the MEMS pitch with respect to the primary was held fixed at the GPI value (18~cm), meaning the size of the pupil in this experiment corresponds to a 5-m telescope.
The comparative mapping between the MEMS actuators and the telescope pupils is shown in Fig.~\ref{fig:experimentscaling}, superimposed over a Kolmogorov phase screen.
The entire square image contains 44x44 actuators, the outer inscribed circle delineates an 8-m pupil (GPI), the inner black square marks 32x32 actuators, and the inner inscribed circle delineates a 5-m pupil (this laboratory experiment).

Measurements were made using the phase-shifting diffraction interferometer\cite{sommargren2002}, a sub-nanometer-absolute-accuracy measuring instrument situated on the ExAO testbed\cite{evans2006a,severson2006} at the LAO.
A spherical wave (f/220) converges from a beam size of 12~mm at the MEMS plane to the surface of a pinhole aligner, where it is combined with a reference beam.
The reference beam emerges from a fiber in spherical waves, emitted through the $2.5$-$\mu$m pinhole to interfere with the reflected test beam.
This reference wave has an excellent wavefront quality ($\lambda/500$) with which to compare the test beam.
The resulting diverging spherical wave fringes are imaged on a CCD camera.
The interferogram is numerically back-propagated using a Huygens method to obtain phase and amplitude data at the MEMS plane.

The free parameters in the saturation experiment were the Fried coherence size, r$_0$, and the woofer pitch, d$_{woof}$.
The procedure was as follows:
An atmospheric turbulence screen with the specified r$_0$ was generated in software by enforcing Kolmogorov turbulence statistics upon an array of random numbers.
Compensation of the atmosphere screen by an ideal woofer DM with the specified d$_{woof}$ was simulated by removing all the power at low spatial frequencies below the appropriate cutoff frequency in the Fourier domain.
The resultant phase screen contained only the higher spatial frequencies.
The MEMS was then ``closed to'' this turbulence screen in closed-loop by using the screen as the reference.
That is, rather than taking a flat shape as the target of the closed-loop algorithm, the target for convergence was the phase screen.
After running 20 closed-loop iterations, the iteration with the lowest rms wavefront error between the target phase screen and the measurement was taken for analysis, along with the corresponding voltage array on the MEMS.
Convergence did not generally improve after 15 iterations.

\begin{figure}[htbp]
\centering\includegraphics[height=11cm,angle=90]{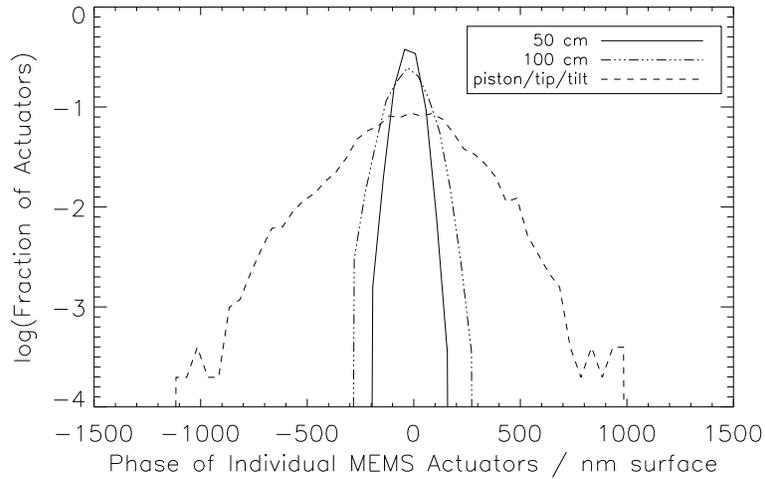}
	\caption{
		\label{fig:individ}
		Histogram of individual actuator phases
		input for the r$_0=15$~cm experiment.
		Dashed line: No woofer (piston and tip/tilt removal only);
		Dashed-dotted line: 100-cm-pitch woofer correction;
		Solid line: 50-cm-pitch woofer correction.
		}
\end{figure}

Fig.~\ref{fig:individ} is a histogram of the stroke inputs at individual actuators' locations.
Three curves are plotted, showing the input Kolmogorov phase screens for r$_0=15~$cm cases with no woofer (piston and tip/tilt removal only), 100-cm-pitch woofer, and 50-cm-pitch woofer.
There were ten screens of each case for these parameters.
This plot shows the stroke range applied to the individual actuators to make the input Kolmogorov phase screens.
The woofer dramatically reduces the phase variance input to the tweeter.

\section{Results}

Taking the best closed-loop iteration in each case, any phase-wrapping (an interferometry artifact) was removed by masking out the bad data points.
Additionally masked out was the 3x3 region centered on a dead actuator located at coordinate [22, 8] (the origin is in the lower-left-hand corner and counting starts with 0).
This was to avoid its being counted as ``saturated'' in each frame.

\begin{figure}[htbp]
\centering\includegraphics[height=11cm,angle=90]{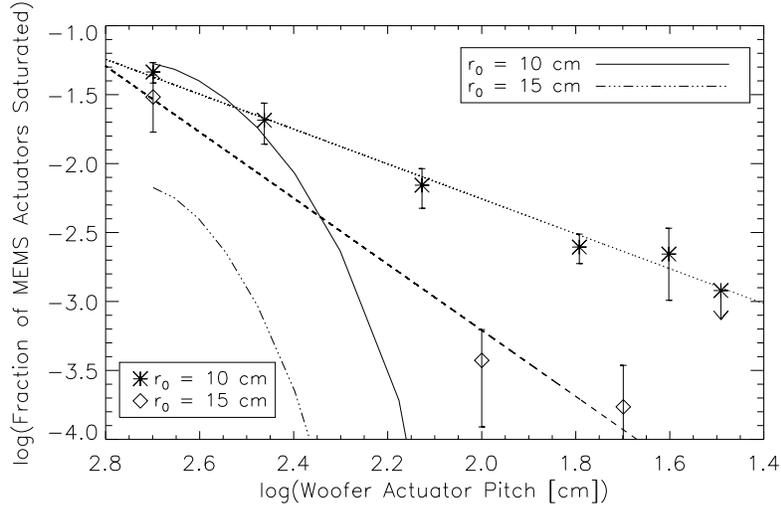}
	\caption{
		\label{fig:results}
		Saturation frequency as a function of woofer pitch.
		Data for experiments at r$_0=10$--15~cm are represented by asterisks and diamonds, respectively,
		while a power-law fit to each experiment is represented by a dotted and a dashed line, respectively.
		Solid and dashed-dotted curves are predictions for r$_0=10$--15~cm, respectively,
		calculated with Equation~\ref{eqn:tweeter} (see details in Section~\ref{sec:analysis}).
		The fitting-error constant $\alpha_F$ used to derive the predictions
		was varied smoothly from 0.14 (piston/tip/tilt only) to 0.28 (continuous facesheet DM)
		over the range d$_{woof}$/D $\textgreater$ 0.25.
		This is due to mapping arguments:
		at the largest d$_{woof}$/D ratios, the pitch of one woofer actuator is a good fraction of the primary diameter,
		and thus whether the woofer actuators overlap or align with the pupil edge becomes important in determining the fitting error.
		This choice (of 4-woofer-actuators-across as the limit for where the DM becomes like a continuous facesheet)
		affects the steepness of the predicted curves but not their endpoints, and we note that only one of our experimental
		data points (290-cm-pitch woofer for r$_0$=15~cm) is well inside this regime where $\alpha_F$ is approximated.
		We see that the saturation incidence if narrow sinc influence functions are assumed
		should have fallen off more steeply than measured.
		Thus decreasing the woofer pitch did not reduce saturation occurrence as much as expected.
		Moderate to high spatial frequencies are saturating more on the MEMS than expected,
		and a simple analytical calculation underpredicts the stroke requirement.
		}
\end{figure}

The corresponding array of voltages applied to the MEMS was used to determine saturation by the criteria that any actuator commanded to the minimum (0~V) or maximum (200~V) was declared ``saturated.''
Table~\ref{tab:woof} lists the set of parameters explored and gives the number of actuators saturated.
The number of actuators tested varies due to different numbers of trials and different numbers of actuators remaining within the unmasked region after discarding phase-wrapped data.
For r$_0=10$~cm and d$_{woof}=31$~cm, zero actuators were saturated out of the 4852 tested.
Therefore, the probability of saturation in that case was calculated using the binomial theorem and taking the outside-3-$\sigma$ probability for the upper limit.
Figure~\ref{fig:results} plots the saturation frequency as a function of woofer pitch for r$_0=10$--15~cm.
Error bars are standard error ($\sigma / \sqrt{N}$).
A power law is fit to each series.

\begin{table}[htbp]
  \centering
  \begin{tabular}{cccc}
    \hline
    r$_{0}$ & Woofer Pitch & \# Actuators & Frac. Saturated Actuators \\
    \textrm{[cm]} & \textrm{[cm]} & Tested & [parts per thousand]\\
    \hline
    10 & 31 & 4852 & $0.01^\dagger$ \\
    10 & 40 & 4359 & 2.2 \\
    10 & 62 & 12~943 & 2.5 \\
    10 & 134 & 6228 & 7.0 \\
    10 & 290 & 2790 & 21 \\
    10 & piston/tip/tilt$^\star$ & 9494 & 46 \\
    \hline
    15 & 50 & 5699 & 0.17 \\
    15 & 100 & 5620 & 0.37 \\
    15 & piston/tip/tilt$^\star$ & 5061 & 30 \\
    \hline
  \end{tabular}
  \caption{
Stroke saturation results:
Fraction of actuators saturated for each set of (r$_0$, d$_{woof}$) parameters, after removing piston and tip/tilt.
The number of actuators tested in each case is equal to the total number of MEMS actuators within the pupil, summed over all trials.
This quantity varies according to two effects: there is a non-uniform number of trials conducted for each set of parameters,
and there is a non-uniform number of actuators analyzed inside the pupil for each trial.
The pupil nominally encircles $\sim600$ actuators, but actuators are masked out and not counted if they are in regions
where phase-wrapping occurs in the interferometric measurement.
  $^\star$At the largest woofer-pitch end, no woofer was used---only piston and tip/tilt were removed over a 5-m aperture.
  We therefore adopted ``500~cm'' as the abscissa for these data
  to be able to plot the corresponding ordinate from column 4 onto Fig.~\ref{fig:results}.
$^\dagger$Upper limit.
}
  \label{tab:woof}
\end{table}

\begin{figure}[htb]
	\centering
  \unitlength1cm
  \begin{minipage}[t]{4cm}
    \begin{picture}(4,4.5)
      \includegraphics[width=4cm,height=4.5cm]{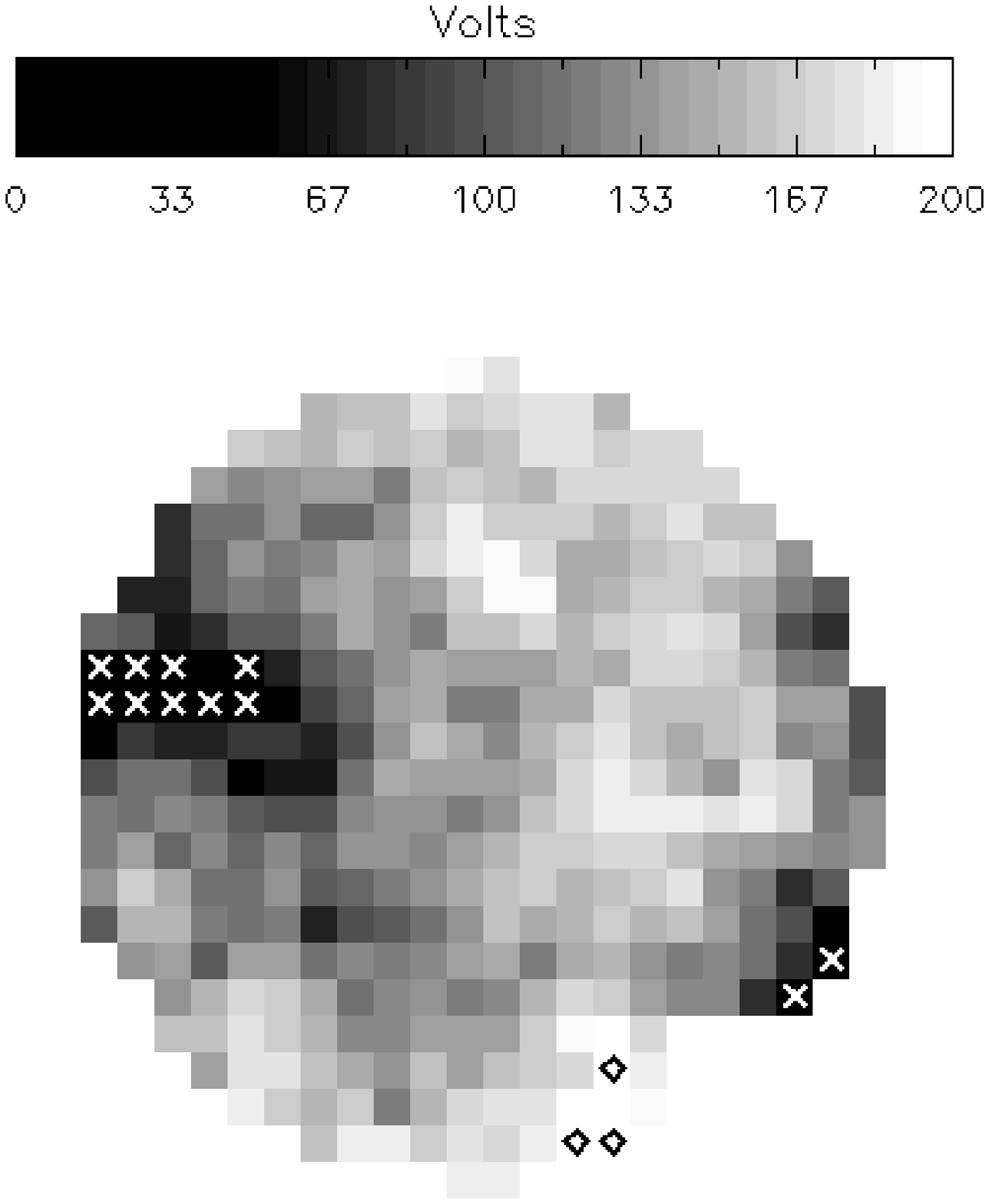}
    \end{picture}
  \end{minipage}
  \begin{minipage}[t]{4cm}
    \begin{picture}(4,4.5)
      \includegraphics[width=4cm,height=4.5cm]{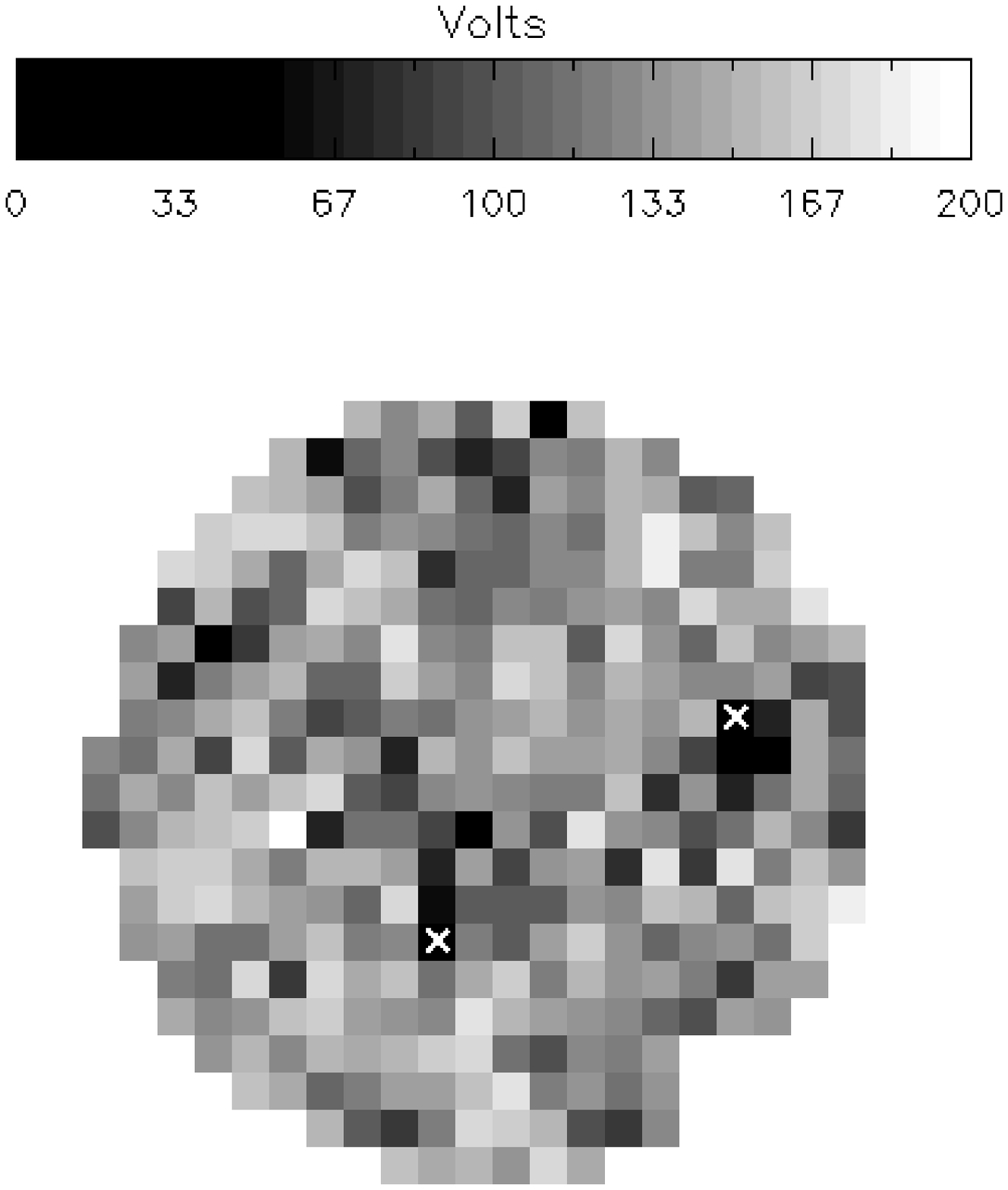}
	\end{picture}
  \end{minipage}
  \begin{minipage}[t]{4cm}
    \begin{picture}(4,4.5)
      \includegraphics[width=4cm,height=4.5cm]{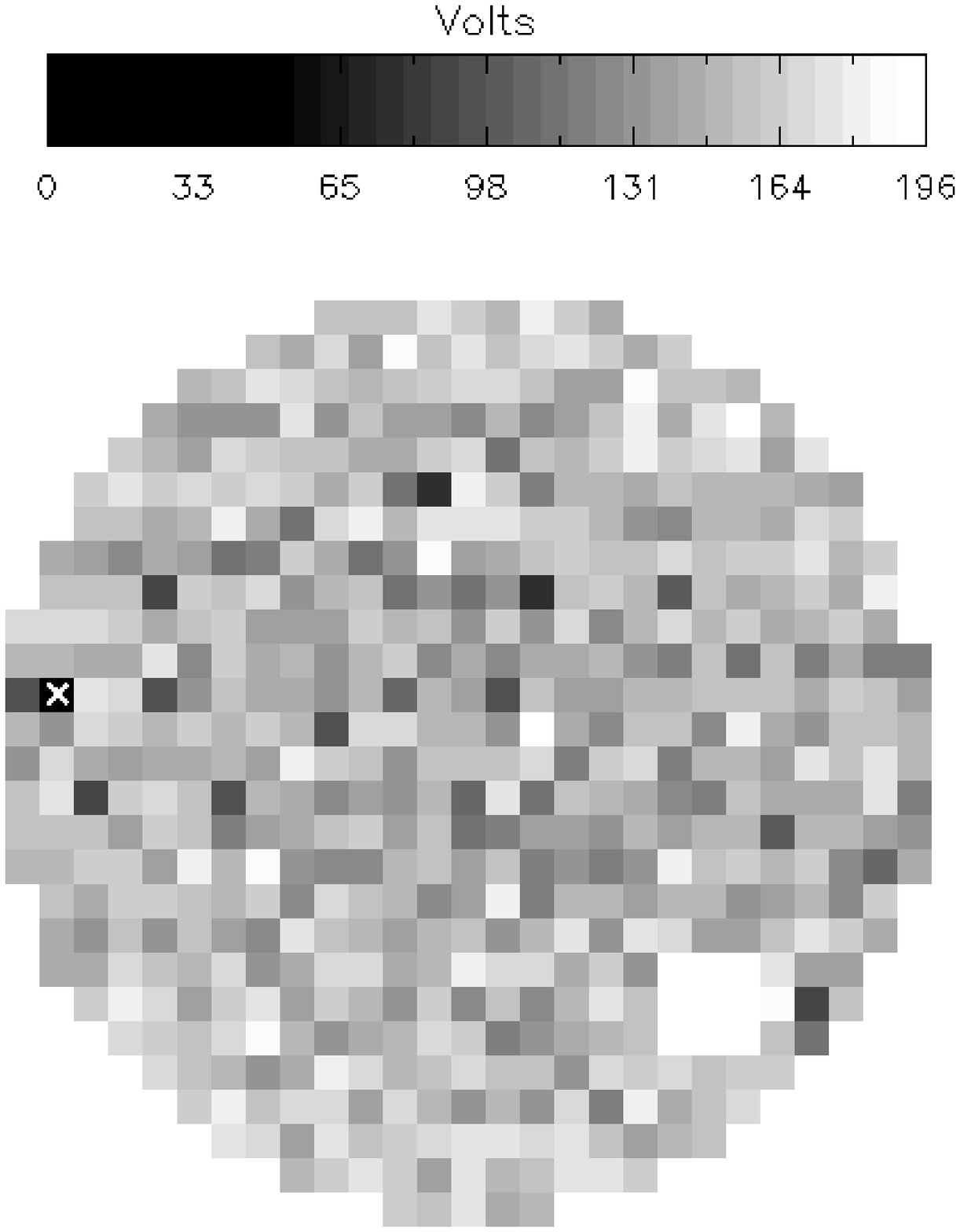}
	\end{picture}
  \end{minipage}
\caption{\label{fig:actsatmaps}
	Maps of MEMS actuator voltages, with saturated actuators marked as follows:
	actuators saturated low are marked with white crosses,
	whereas actuators saturated high are marked with black diamonds.
	Typical results are shown.
	\textit{Left:}
	r$_0=10$~cm, no woofer (piston and tip/tilt removal only).
	Three actuators were saturated high and 11 were saturated low.
	\textit{Center:}
	r$_0=10$~cm, d$_{woof}=134$~cm.
	Two actuators were saturated low (none were saturated high).
	\textit{Right:}
	r$_0=10$~cm, d$_{woof}=62$~cm.
	One actuator was saturated low (none were saturated high).
	}
\end{figure}

Actuator voltage maps give a qualitative picture of close saturation encounters for the MEMS in closing to a Kolmogorov phase screen.
Figure~\ref{fig:actsatmaps} shows three typical actuator voltage maps.
Saturated actuators are identified with a white cross (saturated low) or a black diamond (saturated high).
The first case is a typical iteration for r$_0 = 10~$cm and no woofer (piston and tip/tilt removal only).
The mean number of saturated actuators for these parameters was 16 per trial, and this particular iteration has three actuators saturated high and eleven saturated low.
The second case is a typical iteration for r$_0 = 10~$cm and d$_{woof} = 134~$cm.
The mean number of saturated actuators for these parameters was two per trial, and this particular iteration has two actuators saturated low and none saturated high.
The third case is a typical iteration for r$_0 = 10~$cm and d$_{woof} = 62~$cm.
The mean number of saturated actuators for these parameters was one per trial, and this particular iteration has one actuator saturated low and none saturated high.

Besides using these actuator voltage maps to display the number of saturated actuators, we also notice something qualitative yet important with these plots: saturated actuators occur either singly or else in clumps.
They do not, however, occur in pairs of one up and one down: in the clumpy patches of saturated actuators, all members of the saturated clump are either high or low.
In all, 140 trials total were done (57~046 actuators were tested after masking out bad data), and there are no instances of two adjacent saturated actuators in which one was saturated high and the other low.
Thus, the most extreme case of inter-actuator saturation at opposing positions is not occurring.
However, we will see below that moderate to high spatial frequencies were problematic in another sense.

\begin{figure}[htbp]
\centering\includegraphics[height=11cm,angle=90]{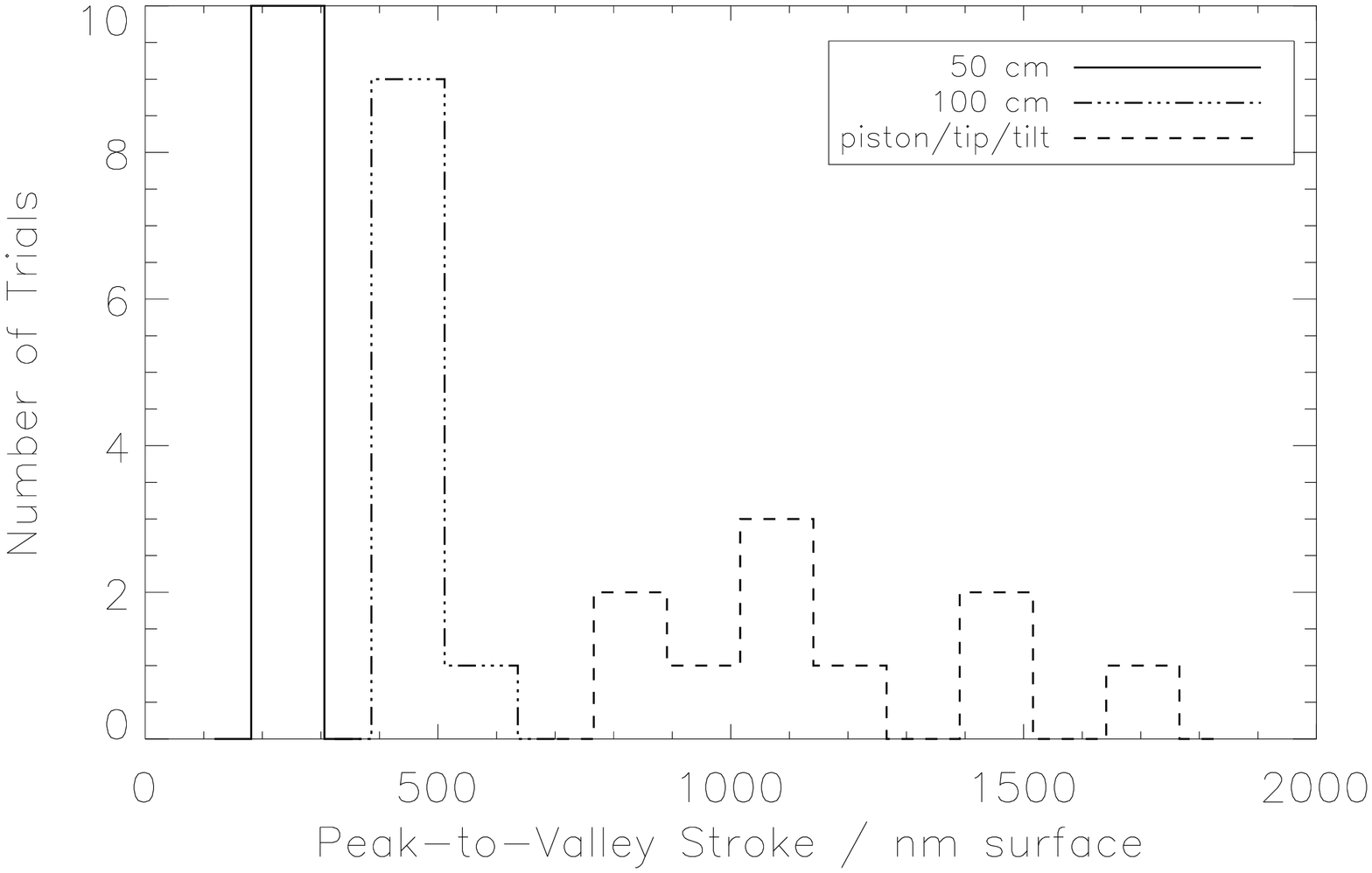}
	\caption{
		\label{fig:pv}
		Histogram of peak-to-valley (i.e. low-spatial-frequency) stroke
		measured for all MEMS actuators in the r$_0=15$~cm experiment.
		Dashed line: No woofer (piston and tip/tilt removal only);
		Dashed-dotted line: 100-cm-pitch woofer correction;
		Solid line: 50-cm-pitch woofer correction.
		The woofer reduces the peak-to-valley stroke on the MEMS by a factor of ~3--4.
		}
\end{figure}

\begin{figure}[htbp]
\centering\includegraphics[height=11cm,angle=90]{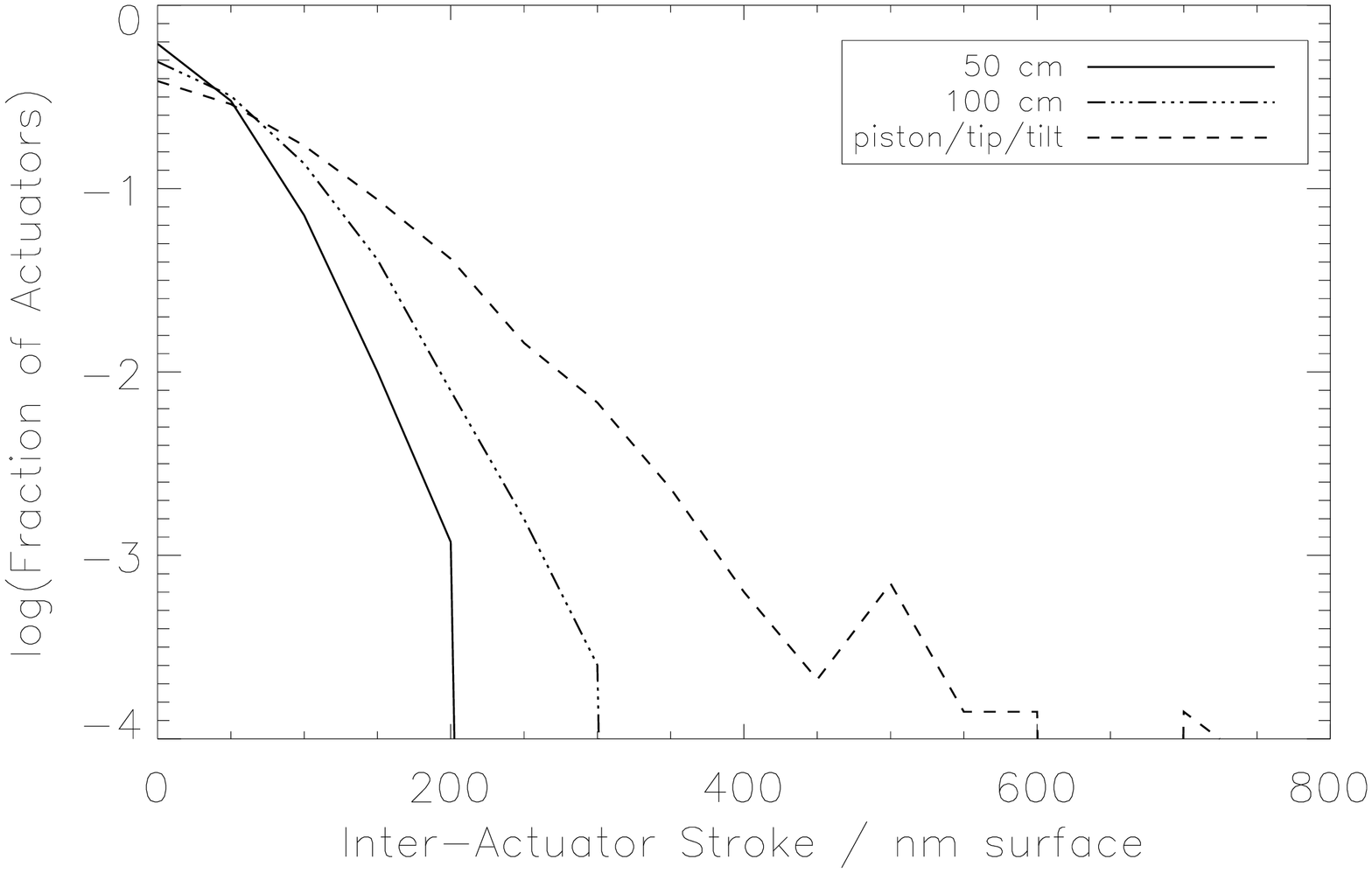}
	\caption{
		\label{fig:ia}
		Histogram of inter-actuator (i.e. high-spatial-frequency) stroke
 		measured for all MEMS actuators in the r$_0=15$~cm experiment.
		Dashed line: No woofer (piston and tip/tilt removal only);
		Dashed-dotted line: 100-cm-pitch woofer correction;
		Solid line: 50-cm-pitch woofer correction.
		The woofer reduces the inter-actuator stroke on the MEMS by a factor of ~2--3.
		}
\end{figure}

Figure~\ref{fig:pv} shows the peak-to-valley stroke histogram for various woofer pitches, as noted in the legend.
The trials without a woofer (piston and tip/tilt removal only) had a peak-to-valley stroke ranging from 0.8 to $1.7~\mu$m surface.
Using a woofer reduces the peak-to-valley stroke required of the MEMS by a factor of ~3--4, down to 0.2 to $0.5~\mu$m surface.
Note that a maximum of $1.5~\mu$m surface stroke was measured for this MEMS, device W107\#X (Table~\ref{tab:spatfreq}), in the low-frequency stroke test.
Obtaining more stroke, up to $1.7~\mu$m surface, with a Kolmogorov phase screen is attributed to the curvature of the inherent shape of the MEMS, which is 450~nm surface for this device.

Figure~\ref{fig:ia} shows the inter-actuator stroke histogram for the same cases.
Inter-actuator stroke is calculated by differencing the measured phase between each pair of neighboring actuators.
Neighbors to a particular actuator are horizontally, vertically, or diagonally adjacent to that actuator.
The woofer reduces the required inter-actuator stroke by a factor of 2--3, a smaller improvement than the peak-to-valley reduction.
This is as expected because the woofer actuator spacing remains much larger than the inter-actuator MEMS spacing ($340~\mu$m), so the inter-actuator stroke is at a higher spatial frequency than the woofer can control.
In addition, a high-amplitude low-frequency aberration will create a slope between two neighboring actuators, introducing further power at the inter-actuator level.

\section{Analysis}\label{sec:analysis}

In order to be able to scale these results to other configurations than our r$_0$ and d$_{woof}$ ranges on a 5m primary, we design a model and test it by using it to predict the measured saturation occurrence.
A histogram of Kolmogorov-turbulence phase (i.e. Fig.~\ref{fig:individ}) follows Gaussian statistics, which can be plotted from theory by calculating the standard deviation with Equation~\ref{eqn:tweeter}.
For a DM having sinc influence functions with the first zero located one actuator away, the stroke of the DM is the same at all spatial frequencies.
Suppose the stroke of such a sinc-influence-function DM is 1.5~$\mu$m surface.
Plot a histogram of Kolmogorov turbulence, and the phase within $+1.5~\mu$m and $-1.5~\mu$m are within the capabilities of this sinc-influence-function DM to correct, whereas the area outside the $\pm1.5~\mu$m cut-off is proportional to the fraction of actuators that saturate.

Table~\ref{tab:spatfreq} and Fig.~\ref{fig:spatfreq} show, however, that the MEMS influence functions are not ideal.
Therefore, instead of using a constant $1.5~\mu$m as the stroke cut-off, we can take the stroke at each spatial frequency, scaling it by a constant multiplier to go from the measurements at 160V to 200V.
Table~\ref{tab:woofgpi} lists the predicted number of actuators saturated for each set of parameters using these methods of finding the area under the Gaussian distribution beyond the cut-off stroke.
Two cut-off strokes are used for comparison, the ideal DM case and the MEMS W107\#X case with stroke varying as a function of spatial frequency.
The measurements from Table~\ref{tab:woof} are repeated for comparison.

\begin{table}[htbp]
  \centering
  \begin{tabular}{ccccc}
    \hline
    r$_{0}$          & Woofer          & Measured Frac. &  Predicted Frac. & Predicted Frac. \\
    \textrm{[cm]} & Pitch              & Sat. Actuators, &  Sat. Actuators, & Sat. Actuators, \\
                      & \textrm{[cm]} & MEMS W107\#X & sinc $1.5\mu$m DM & MEMS W107\#X\\
                           &                        & [parts/thousand]  & [parts/thousand] & [parts/thousand] \\
    \hline
    10 & 31& $0.01^\dagger$ & 0.0 & 0.0 \\
    10 & 40 & 2.2 & 0.0 & 0.0 \\
    10 & 62 & 2.5 & 0.0 & 0.0 \\
    10 & 134 & 7.0 & 0.04 & 0.05 \\
    10 & 290 & 21 & 15 & 13 \\
    10 & piston/tip/tilt$^\star$ & 46 & 53 & 46 \\
    \hline
    15 & 50 & 0.17 & 0.0 & 0.0 \\
    15 & 100 & 0.37 & 0.0 & 0.0 \\
    15 & piston/tip/tilt$^\star$ & 30 & 6.7 & 5.2 \\
    \hline
  \end{tabular}
  \caption{
  Table of measured versus predicted stroke saturation occurrence.
  Predictions were made by using Equation~\ref{eqn:tweeter}
  as the standard deviation to plot a Gaussian distribution
  of all the probable Kolmogorov phase values, taking the area
  under the curve outside the cut-off stroke to be proportional to the fraction of actuators saturated.
  The failure of the predicted values to match the measured values,
  except at the lowest spatial frequencies,
  indicates that the decreasing stroke with increasing spatial frequency of W107\#X
  is producing more saturation than expected for a segmented DM or a DM with sinc influence functions.
  $^\star$At the largest woofer-pitch end, no woofer was used---only piston and tip/tilt were removed over a 5-m aperture.
  We therefore adopted ``500~cm'' as the abscissa for these data
  to be able to plot the corresponding ordinate from columns 3--5 onto Fig.~\ref{fig:results}.
  $^\dagger$Upper limit.
}
  \label{tab:woofgpi}
\end{table}

Saturation was as frequent as expected for cases with larger woofer pitches, but was more frequent than expected for cases with smaller woofer pitches.
The moderate to high spatial frequencies were predicted to cause less saturation on the 1.5-$\mu$m-stroke MEMS than occurred.
Therefore, since the model did not fit the data in this case, the theory could not be used to extrapolate to other configurations.
However, the GPI 4k-MEMS is specified to have 5 times more high-order stroke than the MEMS used in this experiment, which will further reduce stroke saturation.

Surprisingly, even the second attempt at prediction (using a variable stroke cut-off for each particular spatial frequency) did not match the experiments at the moderate to high spatial frequencies.
This can be explained by looking again at Fig.~\ref{fig:spatfreq} and seeing how stroke continues to fall off, albeit less steeply past the knee, no matter what spatial frequency is taken.
If the stroke beyond a certain point were uniform, we would be in the ideal-influence-function regime for those spatial frequencies.
The prediction, by using a single stroke value (at each spatial frequency) for the cut-off of the Gaussian distribution, is still assuming uncorrelated narrow-sinc-function actuators, whereas on the MEMS there are always higher spatial frequencies that limit the stroke.

\section{Discussion}

In systems where saturation cannot be tolerated, the wavefront correctors must have sufficient stroke to correct the amplitude of the phase aberrations.
If deformable mirror stroke is limited, stroke can be conserved for dynamical corrections by
instead using static correctors to remove inherent curvature to the DM itself and for correcting static aberrations in the optical path.
Similarly, the unpowered shape of the DM itself should be as flat as possible so as to conserve stroke for correcting turbulence.
The stroke requirement increases with power in the Kolmogorov spectrum, so better astronomical seeing conditions will give less saturation.
Accordingly, the astronomical site and observing conditions could be favorably selected to minimize stroke saturation. 

In designing a deformable mirror, factors that reduce stroke saturation include decreasing the thickness of the facesheet, increasing the spacing (pitch) between actuators, increasing the operating voltage, and modifying the mechanical design of the actuator (varying parameters such as the capacitor gap, spring constant, flexural rigidity of the plate, or using leverage bending).
These are varied primarily at the research and development phase.

In this work we assume that piston and tip/tilt modes are corrected by other devices, and that the stroke of the woofer DM is high enough to correct the frequencies below the cutoff; therefore, we are only concerned with avoiding saturation on the tweeter DM.
In a dual-DM system, factors that reduce tweeter saturation via sending saturable stroke to a second low-order woofer DM include: employing a woofer with more actuators across the pupil (the subject explored in this paper); and carefully controlling the frequency distribution sent to each wavefront corrector in the control loop\cite{lavigne2008,conan2007}.

Tweeter saturation occurrence can be predicted by analyzing simulations of correcting turbulence screens with model MEMS mirrors\cite{morzinski2007,morzinski2008a,vogel2006,stewart2007}.
Our results show the analytic fitting-error model based on the turbulence power spectrum underpredicts incidence of stroke saturation.
Our original goal in this work was to measure whether stroke saturation would be problematic for GPI.
In order to assure the higher-stroke MEMS mirrors (now being fabricated) are adequate for GPI,
we plan to measure saturation statistics with similar testing on ``engineering-grade'' versions of the 4k-MEMS.


\section{Conclusions}

In this experiment we applied Kolmogorov-turbulence atmospheric phase screens to a MEMS deformable mirror.
The r$_0$ of the phase screens ranged from 10--15 cm, and the pitch of the woofer used to remove the low spatial frequencies ranged from 8--16 actuators across an 8-m primary.

The MEMS when solitary suffered saturation $\sim$4\% of the time.
Using a woofer DM reduces dramatically the amount of phase sent to the MEMS, and thus mitigated MEMS saturation occurrence to a fraction of a percent.

The woofer did not mitigate mid-to-high-spatial-frequency stroke as much as expected.
No neighboring actuators were saturated at opposing positions, meaning the most extreme case of inter-actuator saturation did not occur.

However, moderate to high spatial frequencies did saturate more often than predicted based on a Gaussian distribution of phase and somewhat-idealized influence functions.
This implies that correlations in actuators, i.e. high-spatial-frequency stroke limits, are significant.
Attempts to derive the expected saturation using a simple analytical model underpredicted the stroke requirements, showing that empirical studies are important.


\section*{Acknowledgments}

The authors thank Claire Max, Sandrine Thomas, Andrew Norton, Lisa Poyneer, Jean-Pierre V{\'e}ran, Jean-Fran{\c c}ois Lavigne, and our other lab-mates, co-workers, and colleagues for insightful discussions that contributed to this work.
Furthermore, we thank the anonymous reviewers for their valuable comments that improved the accuracy, clarity, and scope of the paper.

This work was performed under the Michelson Graduate Fellowship for the Jet Propulsion Laboratory, California Institute of Technology, sponsored by the United States Government under a Prime Contract between the California Institute of Technology and NASA.

This research was supported in part by the National Science Foundation Science and Technology Center for Adaptive Optics, managed by the University of California at Santa Cruz under cooperative agreement No.~AST-9876783, PI Claire E.~Max.

Support for this work was also provided by a grant from the Gordon and Betty Moore Foundation to the Regents of the University of California, Santa Cruz, on behalf of the UCO/Lick Laboratory for Adaptive Optics, directed by Don Gavel.  The content of the information does not necessarily reflect the position or the policy of the Gordon and Betty Moore Foundation, and no official endorsement should be inferred.

Portions of this work were performed under the auspices of the U.~S.~Department of Energy by the University of California, Lawrence Livermore National Laboratory under Contract W-7405-ENG-48.
Additionally, support for this work was provided by a minigrant from the Institute of Geophysics and Planetary Physics through Lawrence Livermore National Laboratory.

This research has made use of NASA's Astrophysics Data System Bibliographic Services.

\end{document}